\documentclass[twoside,fleqn]{article}

\usepackage{espcrc2}
\usepackage{epsfig}
\newcommand{\AmS}{{\protect\the\textfont2
  A\kern-.1667em\lower.5ex\hbox{M}\kern-.125emS}}

\begin{document}
 

\title{\vspace{-3.65cm}
       {\normalsize DESY 01-144}    \\[-0.2cm]
       {\normalsize October 2001}   \\
       \vspace{2.72cm}
The spin structure of the $\Lambda$ hyperon in quenched lattice
QCD\thanks{Talk given by M. G\"ockeler at Lat01, Berlin, Germany.}}

\author{
M. G\"ockeler\address{Institut f\"ur Theoretische Physik,
     Universit\"at Regensburg, D-93040 Regensburg, Germany},
R. Horsley\address{John von Neumann--Institut f\"ur Computing NIC,
        D-15735 Zeuthen, Germany},
D. Pleiter$^{\rm b}$,
P.E.L. Rakow$^{\rm a}$,
S. Schaefer$^{\rm a}$,
A. Sch\"afer$^{\rm a}$,
and G.~Schierholz$^{\rm b,}$\address{Deutsches Elektronen-Synchrotron DESY,
      D-22603 Hamburg, Germany} \\
QCDSF Collaboration}
 
\begin{abstract}
It has been suggested to use the production of $\Lambda$ hyperons 
for investigating the nucleon spin structure. The viability of 
this idea depends crucially on the spin structure of the $\Lambda$.
Using nonperturbatively $O(a)$ improved Wilson fermions in the 
quenched approximation we have studied matrix elements of 
two-quark operators in the $\Lambda$. We present results for the axial vector 
current, which give us the contributions of the $u$, $d$, and $s$ 
quarks to the $\Lambda$ spin. 
\end{abstract}
 
\maketitle
 
\section{INTRODUCTION}

In the quark model, the $\Lambda$ hyperon is a $uds$ state, and its spin
is completely carried by the $s$ quark. Combining, on the other hand, the 
experimental results for the $g_1$ structure function of the proton with 
the data on hyperon $\beta$ decay and assuming SU(3) flavour symmetry one
finds that only $\approx 60 \%$ of the $\Lambda$ spin is carried by
$s$ (and $\bar{s}$) quarks, while $u$ ($\bar{u}$) and $d$ ($\bar{d}$)
quarks contribute $\approx - 40 \%$. The spin structure of the $\Lambda$
is of special interest, because its polarisation can easily be 
measured via the self-analysing weak decay $\Lambda \to p \pi^-$.
Indeed, the $\Lambda$ polarisation has been measured
at the $Z^0$ pole in $e^+ e^-$ annihilation~\cite{lep} where it is
mainly due to the $s$ quarks, which when produced via $Z^0$ decays have 
an average polarisation of $-0.91$. Furthermore, $\Lambda$ polarisation
has been studied in deep-inelastic scattering of polarised positrons on 
unpolarised protons in the current fragmentation region, i.e.\ selecting
$\Lambda$'s which most likely originate from the struck quark~\cite{hermes}.
While polarised $s$ quarks are rare in the proton and their contribution 
to electromagnetic processes is suppressed by a factor $Q_s^2 = 1/9$,
there are plenty of polarised $u$ quarks contributing with $Q_u^2 = 4/9$.
Thus even a small correlation between the spins of the $u$'s and 
the $\Lambda$'s could make the $\Lambda$'s useful as probes of the
polarised $u$-quark distribution in the proton~\cite{jaffe}.
Needless to say, however, that the interpretation of the experiments 
is complicated by fragmentation effects, $\Lambda$'s from the decay of 
heavier hyperons, etc.

On the theoretical side, for a spin 1/2 baryon
the fraction $\Delta q$ of the spin 
carried by the quarks (and antiquarks) of flavour $q$ is given
in terms of the forward matrix element of the axial vector current:
\begin{equation}
\langle p,s | \bar{q} \gamma_\mu \gamma_5 q | p,s \rangle
 = 2 \Delta q \cdot s_\mu \,,
\end{equation}
where $s^2 = - m^2$ and the states are normalised according to 
$\langle p,s |p',s' \rangle = 2 E_p (2 \pi)^3 
 \delta(\vec{p} - \vec{p}^{\,\prime}) \delta_{s s'}$.

In the following we shall present preliminary results of a lattice
computation of such matrix elements in the $\Lambda$.

\section{THE SIMULATION}

We have performed quenched simulations with the Wilson gauge action 
at $\beta = 6.0$ 
using nonperturbatively $O(a)$ improved fermions (clover fermions) 
with $c_{\mathrm {SW}} = 1.769$. The lattice size was $16^3 \times 32$.
We worked with nine combinations of hopping parameters:
$\kappa_u = \kappa_d$, $\kappa_s$ $\in \{0.1324, 0.1333, 0.1342 \} $
corresponding to bare quark masses of 
$\approx 166$, $112$, $58$ MeV, respectively. So we can
{\em extra}\hspace{0.5pt}polate 
to the chiral limit in $\kappa_u = \kappa_d$ and 
{\em inter}\hspace{0.8pt}polate 
in $\kappa_s$ to the physical value $\kappa_s^*$.
For the critical hopping parameter $\kappa_c$ we take the value 0.135201
determined from the PCAC quark mass~\cite{dirk}, 
$\kappa_s^* = 0.134138$ was fixed by requiring the pseudoscalar mass 
$m_{\mathrm {PS}}$ for 
$\kappa_u = \kappa_d = \kappa_c$ and $\kappa_s = \kappa_s^*$ to
be equal to the $K^+$ mass $= 494$ MeV (with the scale set by the 
Sommer parameter $r_0 = 0.5$ fm). In the extra- and interpolation 
we assumed a linear dependence of $m_{\mathrm {PS}}^2$ on the quark mass.

As an interpolating field for the $\Lambda$ we used (employing Euclidean
notation from now on)
\begin{equation}
\sum_{x, \, x_4 = t} \epsilon_{i j k} s_i (x) 
\left( u_j^T (x) C \gamma_5  d_k (x) \right)
\end{equation}
with the charge conjugation matrix $C$ ($i$, $j$, $k$: colour indices).
The quark fields were (Jacobi) smeared to improve the overlap with the 
$\Lambda$ state.

As usual, the bare matrix elements are determined from ratios of three-point
functions over two-point functions. In order to reduce the cut-off effects
from $O(a)$ to $O(a^2)$ also in matrix elements, the improvement of the 
fermionic action has to be accompanied by the improvement of the operator
under study. In our case, this is the axial vector current, and for the
renormalised improved operator we can take 
\begin{equation}
A^{\mathrm R}_\mu = Z^0_A (1 + b_A a m) (A_\mu + a c_A \partial_\mu P)
\end{equation}
with the bare axial vector current 
$A_\mu (x) = \bar{q}(x) \gamma_\mu \gamma_5 q(x)$, the pseudoscalar 
density $P(x) = \bar{q}(x) \gamma_5 q(x)$, and the bare quark mass
$am = 1/(2 \kappa) - 1/(2 \kappa_c)$. The improvement term $\partial_\mu P$
vanishes in forward matrix elements, so we do not need the coefficient
$c_A$. For $Z^0_A$ and $b_A$ we take the values from Ref.~\cite{losalamos}:
$Z^0_A = 0.807(2)(8)$, $b_A = 1.28(3)(4)$.

\section{RESULTS}

\begin{figure}
\vspace*{-0.4cm}
\epsfig{file=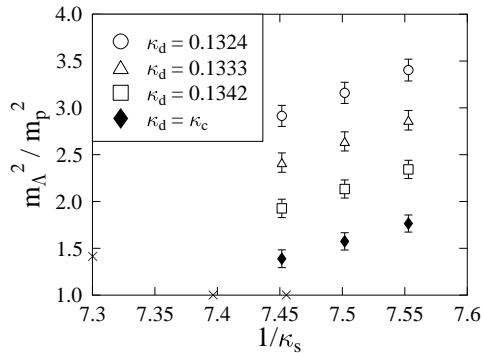,width=7.5cm} 
\vspace*{-1.5cm}
\caption{The square of the ratio $m_\Lambda / m_p$ with the proton mass 
$m_p$ taken in the chiral limit versus $1/\kappa_s$. The different
symbols correspond to the different values of $\kappa_d$ including 
the chiral limit. The crosses (left to right) indicate the physical 
value of $m_\Lambda / m_p$, $1/\kappa_c$, and  $1/\kappa_s^*$, respectively.}
\label{fig.mass}
\end{figure}

From the $\Lambda$ masses at our nine combinations of $\kappa_d$, 
$\kappa_s$ we have computed $\Lambda$ masses at $\kappa_d = \kappa_c$
by linear extrapolation of $m_\Lambda^2$ in $1/\kappa_d$.
These 12 masses are plotted in Fig.\ref{fig.mass}. Our
value for $\kappa_s^*$ reproduces quite accurately the ratio 
$m_\Lambda / m_p = 1.19$.
This fits in nicely with the observation that $m_\Sigma / m_p$ and
$m_\Xi / m_p$ are also rather well reproduced by quenched simulations
(see, e.g., \cite{dirk}). The plot shows clearly the breaking of 
the SU(3) flavour symmetry.
In particular, the dependence of the masses on $\kappa_d$ is rather 
pronounced. 

\begin{figure}
\vspace*{-0.4cm}
\epsfig{file=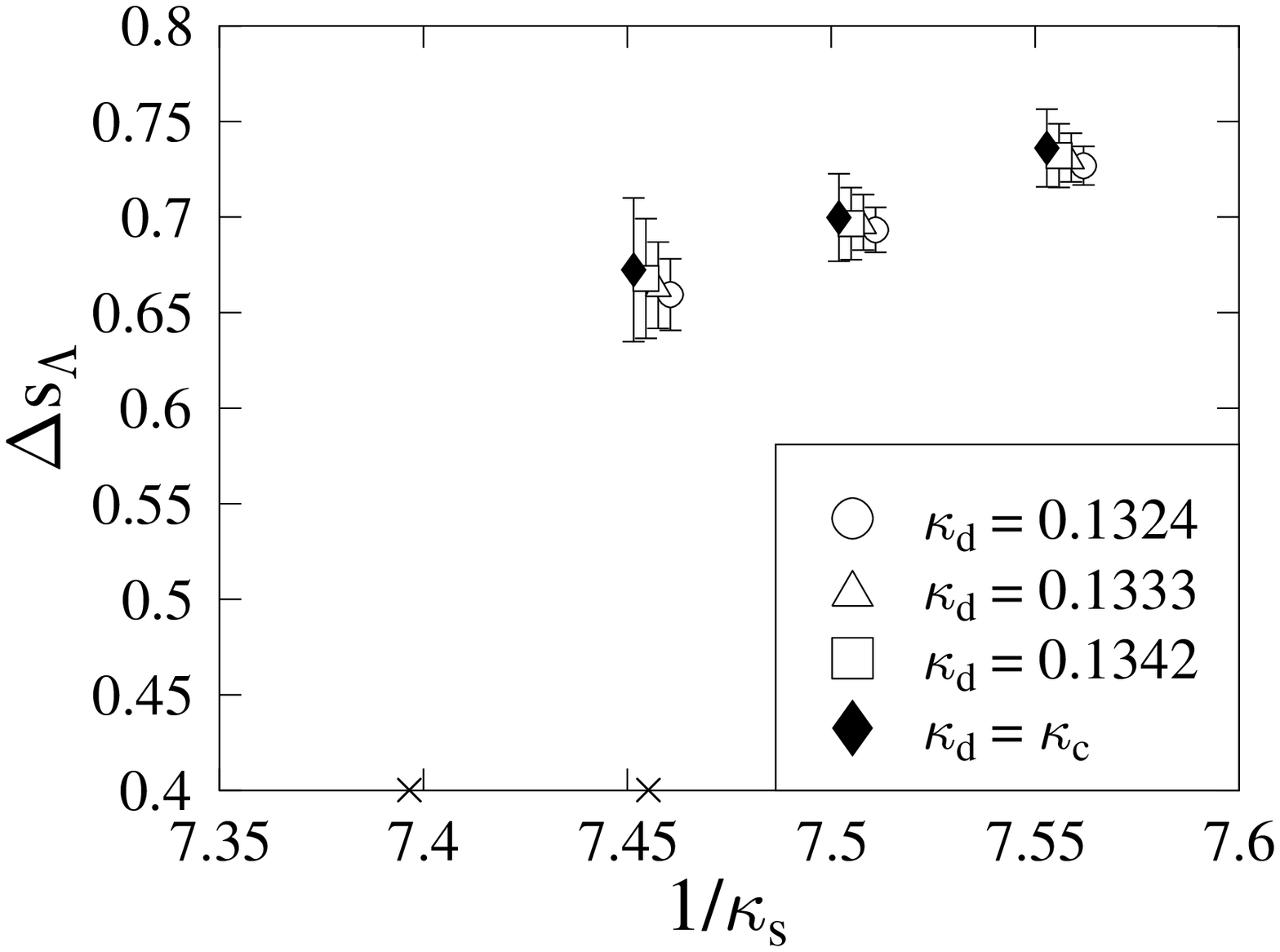,width=7.5cm} \\[-0.4cm]
\epsfig{file=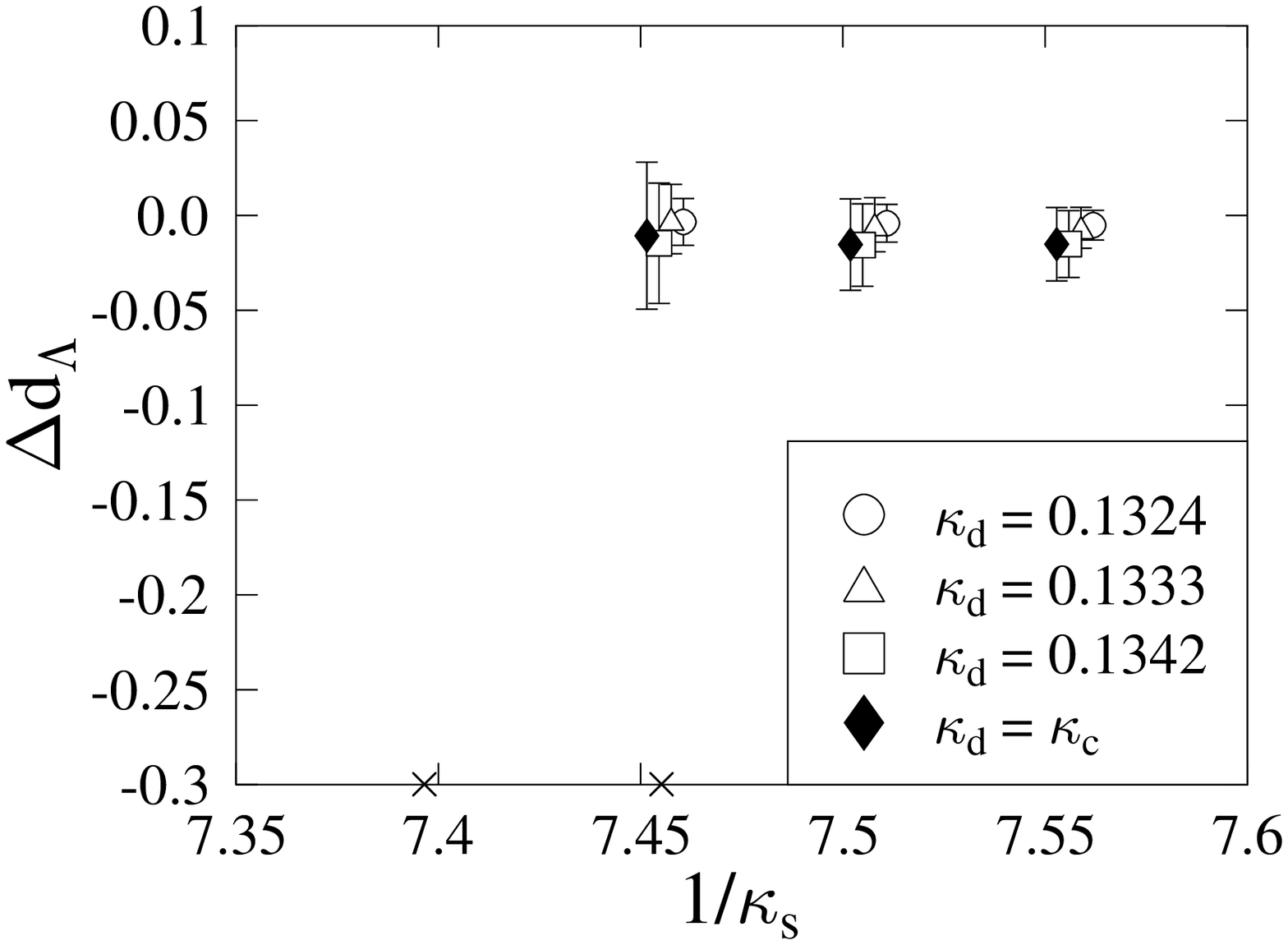,width=7.5cm} 
\vspace*{-1.5cm}
\caption{$\Delta s$ and $\Delta d$ in the $\Lambda$ versus $1/\kappa_s$.
The results for the different values of $\kappa_d \neq \kappa_c$
have been displaced horizontally.
The crosses indicate $1/\kappa_c$ and  $1/\kappa_s^*$.}
\label{fig.delta}
\end{figure}

In Fig.\ref{fig.delta} we plot our (preliminary)
results for $\Delta s$ and $\Delta d$ in the $\Lambda$ versus $1/\kappa_s$.
In contrast with the case of $m_\Lambda / m_p$, the dependence on $\kappa_d$
is rather weak. The values corresponding to $\kappa_d = \kappa_c$ have been
obtained by extrapolating the bare matrix elements linearly in $1/\kappa_d$.
Interpolating the bare matrix elements for $\kappa_d = \kappa_c$
linearly in $1/\kappa_s$ to $1/\kappa_s^*$ we obtain the desired 
$\Lambda$ matrix elements given in the last line of the following
table:

\begin{center}
\renewcommand{\arraystretch}{1.4}
  \begin{tabular}[h]{lll} \hline 
     {}  & \multicolumn{1}{c}{$\Delta u_\Lambda = \Delta d_\Lambda$} & 
    \multicolumn{1}{c}{$\Delta s_\Lambda$}\\ \hline 
    quark model & 
   \multicolumn{1}{c}{$0$}         & \multicolumn{1}{c}{$1$}      \\
    exp. + SU(3)$_{\mathrm F}$ & 
          $-0.17(3)$                    &          $0.63(3)$           \\
    MC + SU(3)$_{\mathrm F}$ &
            $-0.016(9)$               &         $0.65(2)$               \\
    this work   &
                 $-0.01(4)$       &        $0.67(3)$                 \\
   \hline
  \end{tabular}
\end{center}

In the third line we give the $\Lambda$ matrix elements as they follow from
our Monte Carlo results for the {\em proton} matrix elements by the use of 
SU(3)$_{\mathrm F}$. They agree quite well with the matrix elements 
computed directly. Since the flavour symmetry breaking effects in the
matrix elements are rather small (see Fig.~\ref{fig.delta}), this is
hardly surprising.
The second line contains the values of the $\Lambda$ matrix elements 
computed from the proton spin structure under the assumption of flavour
SU(3)(see, e.g., \cite{ashery}). All these results differ markedly from 
the predictions of the (naive) quark model shown in the first line.

Concerning the Monte Carlo results one should keep in mind that they
were obtained at a fixed value of the lattice spacing. So a continuum
extrapolation is not possible. Furthermore we used the quenched 
approximation and neglected quark-line disconnected contributions. 
Therefore it might be more reasonable to compare with estimates of
the valence quark contribution, for which, e.g., Ashery and 
Lipkin~\cite{ashery} find 
$ \Delta u_\Lambda = \Delta d_\Lambda = -0.07(4)$, 
$ \Delta s_\Lambda = 0.73(4)$.
Compared with the numbers from the second line of our table,
$\Delta d_\Lambda$ gets closer to our Monte Carlo result, 
while $\Delta s_\Lambda$ becomes somewhat larger. 

\section{SUMMARY}

Our Monte Carlo investigation of the spin structure of the $\Lambda$ 
hyperon leads to satisfactory agreement with the phenomenological numbers,
although the comparison is not straightforward due to the use of the 
quenched approximation. SU(3) flavour symmetry appears to be much
less violated in the matrix elements of the axial vector current than in
the baryon masses, in agreement with the empirical
observation that the hyperon semileptonic
decays can be parametrised rather well assuming SU(3) flavour symmetry. 
The relevance of our findings for using 
$\Lambda$'s as a probe of the nucleon spin structure in 
deep-inelastic lepton-nucleon scattering remains to be studied.
But we expect them to shed some light also on the lattice results for
the nucleon spin structure.

\section*{ACKNOWLEDGEMENTS}
This work is supported by the DFG (Schwer\-punkt ``Elektromagnetische Sonden'')
and by BMBF. The numerical calculations were performed on the Quadrics 
computers at DESY Zeuthen. We wish to thank the operating staff 
for their support.


\end{document}